 \documentclass[final,5p,times,twocolumn]{elsarticle}

\usepackage{amssymb}
\usepackage{amsmath}
\usepackage{lipsum}
\usepackage[utf8]{inputenc}
\usepackage[T1]{fontenc}
\usepackage[spanish,catalan,english]{babel}
\usepackage{todonotes}
\usepackage{hyperref}

\journal{Physics Letters B}

\newcommand{\nestor}[1]{#1}
\newcommand{\victor}[1]{#1}
\newcommand{\elena}[1]{#1}

\begin{document}

\begin{frontmatter}



\title{Lindblad-driven recombination of the \texorpdfstring{$X(3872)$}{X(3872)} tetraquark}


\author[first]{N\'estor Armesto}
\author[second]{Miguel Ángel Escobedo}
\author[first]{Elena G. Ferreiro}
\author[first]{Víctor López-Pardo}
\affiliation[first]{organization={Instituto Galego de Física de Altas Enerxías IGFAE, Universidade de Santiago de Compostela},
            addressline={Rúa de Xoaquín Díaz de Rábago, s/n}, 
            city={Santiago de Compostela},
            postcode={15782}, 
            state={Galicia},
            country={Spain}}
\affiliation[second]{organization={Departament de Física Quàntica i Astrofísica and Institut de Ciències del Cosmos, Universitat de Barcelona},
            addressline={Martí i Franquès 1}, 
            city={Barcelona},
            postcode={08028}, 
            state={Catalonia},
            country={Spain}}
\begin{abstract}
The internal structure of the exotic meson $X(3872)$ remains an open question. We investigate its production in heavy-ion collisions under the hypothesis that it is a compact tetraquark. To this end, we derive a coalescence model from the Lindblad equation, assuming that unbound heavy quarks are thermalized within the quark-gluon plasma and that the adiabatic approximation holds. Using this model, we predict the nuclear modification factor of the $X(3872)$ at LHC energies, with proton-proton baseline cross sections estimated from available experimental data. 
We also consider the effect of simplifying assumptions on the model, and a complementary approach based on chemical equilibration.
Our results indicate that recombination is the dominant production mechanism for a tetraquark $X(3872)$. It leads to a significant yield enhancement in heavy-ion collisions, and suggests that the nuclear modification factor is a powerful observable for probing the exotic nature of this state.
\end{abstract}



\begin{keyword}
Exotic states \sep coalescence \sep heavy-ion collisions \sep finite temperature



\end{keyword}

\end{frontmatter}




\section{Introduction}
\label{introduction}
The landscape of hadron spectroscopy has been reshaped over the last two decades by the discovery of numerous exotic states in the heavy quark sector~\cite{Brambilla:2022ura}. These particles, which are difficult to classify within the conventional quark model, have spurred competing theoretical descriptions of their nature, most notably as either compact multi-quark states (e.g., tetraquarks) or as loosely-bound hadronic molecules. Among these, the $X(3872)$, also denoted as $\chi_{c1}(3872)$, holds a special status, being the first such charmonium-like state to be observed~\cite{Belle:2003nnu}. Despite years of study, its internal structure remains an unresolved question in Quantum Chromodynamics (QCD).

A novel environment for probing the nature of the $X(3872)$ has emerged with its recent detection in ultrarelativistic heavy-ion collisions at the LHC~\cite{CMS:2021znk}. These collisions create a deconfined medium of quarks and gluons (QGP), where hadron properties are modified. The study of conventional quarkonium in this medium has long served as a fundamental tool for diagnosing the properties of the QGP itself, primarily through the mechanism of color screening~\cite{Andronic:2015wma}. By extending this paradigm to exotic states, a new window of investigation is opened. The behavior of the $X(3872)$ within the QGP is expected to be highly sensitive to its configuration; its production and survival are powerful observables that can help distinguish between a compact tetraquark and an extended molecular structure, thereby providing crucial insights into both the properties of the hot medium and the fundamental nature of exotic hadrons~\cite{Wu:2020zbx,Chen:2021akx,Yun:2022evm,MartinezTorres:2014son}.

In~\cite{Armesto:2024zad}, we developed a potential model for the $X(3872)$ at finite temperature, assuming it is a compact tetraquark. This allowed us to study its suppression in the QGP. However, we only took into account the dissociation of an initially produced $X(3872)$, neglecting the possibility of recombination from unbound heavy quarks and antiquarks in the medium.
Recent results from the CMS Collaboration~\cite{CMS:2021znk} have shown that the $\frac{X(3872)}{\psi(2S)}$ production ratio is higher in Pb–Pb collisions than in proton-proton collisions, suggesting that regeneration mechanisms involving coalescence and recombination play a significant role in the formation of the $X(3872)$. 
This enhancement cannot be explained by suppression models and requires a framework that consistently incorporates both dissociation and recombination dynamics.
In this work, we address this gap by deriving a coalescence model for the $X(3872)$ from the Lindblad equation, under the assumptions that unbound heavy quarks are thermalized within the QGP and that the adiabatic approximation holds. This framework enables us to predict the nuclear modification factor, $R_{AA}$, of the $X(3872)$ at LHC energies, using proton-proton baseline cross sections estimated from available experimental data.

Our findings suggest that if the $X(3872)$ is indeed a compact tetraquark, recombination would be the dominant production mechanism in heavy-ion collisions. 
Like in our previous work~\cite{Armesto:2024zad}, our aim is to provide order-of-magnitude estimates and to highlight the potential of the $R_{AA}$ as a discriminating observable for the exotic nature of the $X(3872)$. Our results should be confronted with future studies assuming a molecular structure for the $X(3872)$, as well as with forthcoming experimental measurements in heavy-ion collisions.

Moreover, in order to confirm our results, we have also studied the case in which chemical equilibration is reached for the $X(3872)$ in the QGP. In this case, recombination is significantly smaller but of the same order of magnitude as with our coalescence model. The difference can be tracked back to the fact that, in a Bjorken expanding medium, recombination of the $X(3872)$ tends to happen at earlier times when the medium volume is smaller and the temperature larger. Contrary to this, in the chemical equilibration scenario, the $X(3872)$ yield is fixed at the hadronization time, when the volume is larger and the temperature smaller. 

The paper is organized as follows: In Section~\ref{sec:potential}, we recall our baseline potential model used to describe the $X(3872)$. In Section~\ref{sec:Lindblad}, we derive the coalescence model starting from the Lindblad equation. Next, we discuss our phenomenological inputs and present our predictions for the $R_{AA}$ of the $X(3872)$ in heavy-ion collisions. In Section~\ref{sec:chemical}, we present a complementary approach based on the assumption of chemical equilibration
of any bound quarkonium-like state. In Section~\ref{sec:gaussian}, a \nestor{rough, } qualitative discussion is made on the possible extension of the model for a hadronic molecule, based on a Gaussian approximation for the wave function.
Finally, we summarize our findings and discuss their implications in Section~\ref{sec:summary}.

\section{Baseline potential model}
\label{sec:potential}

In Ref.~\cite{Armesto:2024zad}, the $X(3872)$ was described as a compact tetraquark bound state
within the Born-Oppenheimer approximation, where the heavy charm quarks move slowly in the
potential generated by light degrees of freedom.  
This potential, extracted from lattice-inspired hybrid calculations~\cite{Capitani:2018rox}, provides the foundation for
the in-medium description developed here.

The in-medium potential is written as
\begin{equation}
V(r,T) = \mathrm{Re}\,V(r,m_D(T)) + i\,\mathrm{Im}\,V(r,m_D(T)),
\label{eq:potential}
\end{equation}
where $r$ is the heavy quark-antiquark separation and $m_D(T)$ the Debye screening mass that encodes
color screening in the QGP.

The real part of the in-medium potential is obtained by applying a linear-response
(polarization/permittivity) procedure to the vacuum Born-Oppenheimer
potential used for the tetraquark.  We start from the lattice-inspired vacuum parametrization
\[
V_{\mathrm{vac}}(r) = \frac{A_{-1}}{r} + A_0 + A_2 r^2,
\]
and obtain the in-medium potential using the HTL permittivity \(\varepsilon(p,m_D)\)
according to \(V(p)=V_{\mathrm{vac}}(p)/\varepsilon(p,m_D)\).  Following the
method of~\cite{Burnier:2015nsa,Burnier:2015tda,Lafferty:2019jpr} and enforcing smooth
matching to the vacuum potential for \(m_D\to 0\), one obtains the form used
in \cite{Armesto:2024zad}:
\[
\begin{aligned}
\mathrm{Re}\,V(r,m_D) &= A_{-1}\!\left(m_D+\frac{e^{-m_D r}}{r}\right) + A_0 \\
&\quad + A_2\!\left[\frac{6}{m_D^2}\big(1-e^{-m_D r}\big)
- \Big(2r^2 + \frac{6r}{m_D}\Big)e^{-m_D r}\right].
\end{aligned}
\]
This expression reproduces the vacuum form for \(m_D\to 0\) at short distances,
is screened at finite \(m_D\), and saturates at large \(r\) to
\(\;A_{-1}m_D + A_0 + 6A_2/m_D^2\;\).  

The imaginary part of the potential arises from Landau damping and the scattering of the
heavy quark with medium gluons, which induces a finite in-medium width.
It is approximated by
a temperature-dependent $r$-constant $\mathrm{Im}\,V(r,T) = -\Gamma(T)/2$,
with $\Gamma(T)$ the thermal decay width modeled as
\begin{equation}
\Gamma(T) = \nestor{2}A_{-1}T + \nestor{2}A_2\,\frac{T}{m_D^3},
\label{eq:GammaT}
\end{equation}
with parameters $A_{-1}=18.904~\mathrm{MeV\cdot fm}$ and
\mbox{$A_2=398.666~\mathrm{MeV\cdot fm^{-2}}$}, the same constants that appear in the real part of
the potential.

The rationale for treating the imaginary part as a constant is as follows. A heavy quark-antiquark pair within a tetraquark typically exists in an octet configuration. In the $r\to 0$ limit, this octet is perceived by the medium as a heavy adjoint color source. Conversely, in the $r\to\infty$ limit, the pair behaves as two uncorrelated heavy quarks. Notably, in the large $N_c$ limit, the decay width of both systems is identical. This suggests that the imaginary part of the potential in an octet configuration depends only weakly on $r$, a behavior that is consistent with perturbative computations.

The dissociation temperature is obtained by solving the Schr$\ddot{ \rm o}$dinger equation using the complex
potential. Since the imaginary part is a constant, it does not affect the solution. The bound solution disappears at a dissociation temperature
$T_d\simeq 350~\mathrm{MeV}$,
defining the point where Debye screening overcomes the confining potential.  
Below $T_d$, the real part of the potential still supports a shallow bound state,
but the imaginary part leads to a finite width that grows with temperature.
The width determines the
survival probability of an initially produced $X(3872)$ as the medium expands and cools.

Assuming a Bjorken expansion, $T(t)=T_0(t_0/t)^{1/3}$,
the survival probability from the initial time $t_0$ to a later time $t$ is
\begin{equation}
S(t) = \exp\!\left[-\!\int_{t_0}^{t} \Gamma(T(\tau))\, d\tau\right].
\label{eq:survival}
\end{equation}
Using the parameters $T_0=500~\mathrm{MeV}$ and $t_0=0.6~\mathrm{fm}$, which correspond to commonly used values at
LHC energies, and stopping the evolution at the time in which $T_c = 180$ MeV, around the phase transition value,
one obtains $S(t)\simeq0.1$-$0.3$ at freeze-out for Pb-Pb collisions.

\begin{figure}[ht]
\centering
\includegraphics[width=0.48\textwidth]{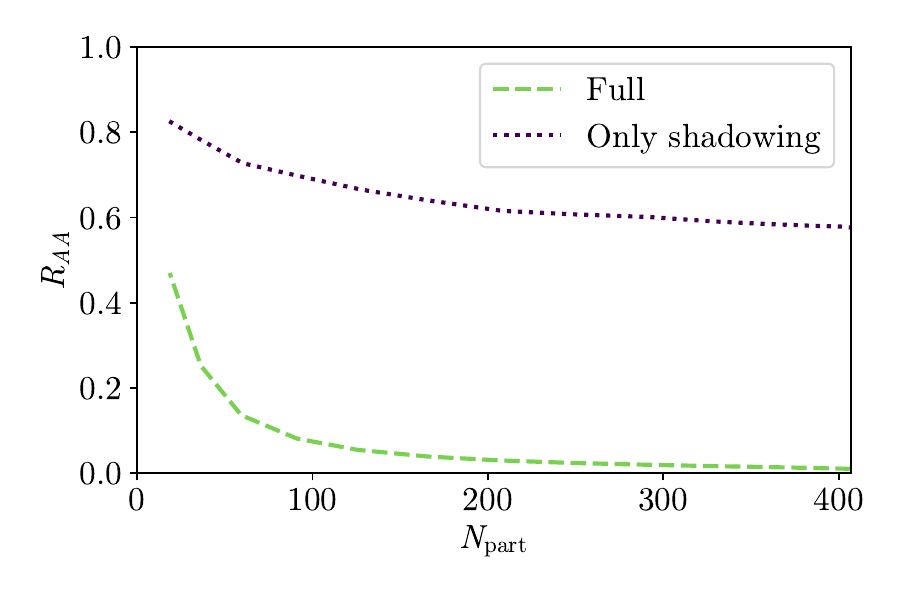}
\caption{Prediction for $R_{AA}$ of $X(3872)$ in Pb-Pb collisions at LHCb conditions. The dotted line considers only cold nuclear matter effects (shadowing) following the model discussed in~\cite{Escobedo:2021ifp}, while the dashed line also includes dissociation.}
\label{fig:RAAs}
\end{figure}

Moreover, using Equation~\eqref{eq:survival}   one can compute the nuclear modification factor, $R_{AA}$, once we know how the temperature seen by the bound state changes with time and position. The initial temperature at a given point is computed using the model developed in~\cite{Escobedo:2021ifp} that takes into account cold nuclear matter effects in the form of shadowing of parton densities inside nuclei. If the initial temperature is larger than $T_d$, then the state is not formed. Otherwise, we compute the survival probability. The results obtained following this procedure can be seen in Figure~\ref{fig:RAAs}, where we have considered the range in transverse momentum and pseudo-rapidity that LHCb can cover for this observable, $0.5~\mathrm{GeV}/c < p_T < 50~\mathrm{GeV}/c $  and $ 2 < \eta < 5$.
We observe a strong suppression and only a relatively mild influence of cold nuclear matter effects. We stress that, for the moment, we have not yet considered recombination effects.

The prompt production of the $X(3872)$ in Pb-Pb collisions has been recently 
reported by the CMS Collaboration in~\cite{CMS:2021znk}, where the 
yield ratio relative to $\psi(2S)$ was found to be
\[
\rho^{\mathrm{PbPb}} \equiv 
\left(\frac{N_{X(3872)}}{N_{\psi(2S)}}\right)_{\!\mathrm{PbPb}} 
= 1.08 \pm 0.49_{\mathrm{(stat)}} \pm 0.52_{\mathrm{(syst)}}
\]
in the kinematic window $|y|<1.6$ and $15 < p_T < 50~\mathrm{GeV}/c$. 
For comparison, the corresponding ratio in proton-proton collisions is approximately 
$\rho^{\mathrm{pp}} \simeq 0.10$ in the same kinematic region.
From these quantities one can define the double ratio
\begin{equation}
R_{\mathrm{double}} \;=\;
\frac{(N_{X(3872)}/N_{\psi(2S)})_{\mathrm{PbPb}}}
{(N_{X(3872)}/N_{\psi(2S)})_{\mathrm{pp}}}
\;=\; \frac{\rho^{\mathrm{PbPb}}}{\rho^{\mathrm{pp}}}
\;=\; 10.8 \pm 7.5,
\label{eq:Rdouble}
\end{equation}
where the uncertainty is obtained by combining statistical and systematic 
errors in quadrature.  

This double ratio connects the nuclear modification factors of the two states,

\[
\begin{aligned}
\frac{R_{AA}(X(3872))}{R_{AA}(\psi(2S))} = R_{\mathrm{double}},
\\
R_{AA}(X(3872)) = R_{\mathrm{double}} \times R_{AA}(\psi(2S)).
\label{eq:raa_relation}
\end{aligned}
\]

The nuclear modification factor of the $\psi(2S)$ has been measured by CMS in 
Pb-Pb collisions at $\sqrt{s_{NN}}=5.02$~TeV~\cite{CMS:2017uuv}, 
with values exceeding $0.1$ in the high-$p_T$ region. Adopting a representative value consistent with those data,
$R_{AA}(\psi(2S)) = 0.15 \pm 0.075$,
we obtain
$R_{AA}(X(3872))=1.62 \pm 1.38$.
Although the inferred value carries large uncertainty,
it highlights the intriguing possibility that the production of the 
$X(3872)$ in heavy-ion collisions may be enhanced relative to that in proton-proton collisions, consistent with the 
expectation from charm-quark recombination.

In the following Section, we extend the baseline potential model
of Ref.~\cite{Armesto:2024zad} by introducing coalescence\footnote{We use coalescence and recombination as synonymous.} contributions implemented in two
complementary approaches:  a non-equilibrium kinetic treatment and a chemical-equilibrium limit.

\section{Recombination model from the Lindblad equation}
\label{sec:Lindblad}
Let us assume that the $X(3872)$ is a compact tetraquark made of a heavy quark-antiquark pair ($c\bar{c}$) in an octet configuration plus a light quark-antiquark pair ($q\bar{q}$) also in an octet configuration. Assuming that the heavy quarks move non-relativistically around the center-of-mass of the tetraquark and that $\Lambda_{QCD}$ is larger than the binding energy, we can describe the evolution using the Born-Oppenheimer approximation. This leads to a potential model where the potential is given by the energy of the light degrees of freedom in the presence of static color sources (the heavy quarks)~\cite{Soto:2020xpm}. At finite temperature, this potential gets modified due to the presence of the medium~\cite{Armesto:2024zad}. However, as in the case of conventional quarkonium~\cite{Brambilla:2017zei}, this potential only describes the evolution of the time-ordered correlator of the heavy quark-antiquark pair. If we wish to describe the evolution of the number of states within the medium we have to consider the evolution of the density matrix. This can be done treating quarkonium as an open quantum system~\cite{Akamatsu:2011se}. In the case $T\gg E$, where $T$ is the temperature of the medium and $E$ the binding energy of the state, the evolution of the density matrix is given by a Lindblad equation~\cite{Brambilla:2017zei,Brambilla:2016wgg}:
\begin{equation}
\label{eq:LindbladOrig}
    \frac{d\rho}{dt} = -i[H,\rho] + \sum_n \left(C_n \rho C_n^\dagger - \frac{1}{2}\{C_n^\dagger C_n, \rho\}\right)\,,
\end{equation}
where $H$ is the Hamiltonian of the system and the $C_n$ are the collapse operators that describe the interaction with the medium. We can also define an effective non-Hermitian Hamiltonian as
\begin{equation}
    H_{\rm eff} = H - \frac{i}{2}\sum_n C_n^\dagger C_n\,.
\end{equation}
In our case, this effective Hamiltonian consists in the sum of the standard kinetic term plus the complex potential described in~\cite{Armesto:2024zad}. This implies that $\mathrm{Im} V(r)=-\frac{1}{2}\sum_n C_n^\dagger C_n$. 
This Lindblad equation 
describes the evolution of a single heavy quark-antiquark pair in the octet configuration. 

The key assumptions of the model are the following:
\begin{itemize}
\item Validity of the Lindblad equation:
  The interaction of a heavy $c\bar c$ pair with the surrounding medium is
  described by a Markovian Lindblad master equation, in which the environment
  induces decoherence and stochastic transitions between color and bound-state
  configurations.

\item Dilute heavy-quark limit:
  It is a rare event that two heavy particles are close enough in configuration
  space to interact.  Accordingly, $n$-body interactions involving multiple heavy
  quarks or antiquarks are neglected, and only binary correlations are retained.

\item Molecular chaos hypothesis:
  Unbound heavy quarks are approximated by uncorrelated particles
  that are locally thermalized with the medium.  

\item Adiabatic evolution:
  The time dependence of the effective non-Hermitian Hamiltonian and operators entering the
  Lindblad equation is slow compared with the intrinsic time scales of the bound
  states.  The adiabatic theorem can therefore be applied to propagate the bound
  component of the density matrix, and to define a survival probability determined
  by the imaginary part of the potential and a recombination rate.

\item Rare recombination approximation:
  The probability of forming an $X(3872)$ bound state through recombination is
  small compared with the number of available unbound pairs.  Hence, the overall
  population of free charm quarks remains effectively constant during the
  evolution, and backreaction effects on the unbound sector can be neglected.
  
\item Later on, we are going to define what we call the target Wigner distribution. This is the Wigner distribution of the bound state dressed by the collapse operators. We assume that the target Wigner distribution is equal to the bound state Wigner distribution. The rationale for this approximation is the following. The Wigner distribution is a function that only has support over a narrow region whose size is related with the size of the bound state. Any dressing of this distribution should not change this support sizably. Therefore, any dressing that keeps the normalization will not change our results qualitatively. At the moment, this is enough for our purposes. If, in the future, a microscopic model able to produce the value of the collapse operators is developed, then it is straight-forward to compute the target Wigner distribution.
\end{itemize}

Under these hypotheses, the dynamics of the heavy-quark sector can be reduced
to an effective open quantum system in which the evolution of the reduced density
matrix is driven by the non-Hermitian Hamiltonian and stochastic jumps encoded
in the Lindblad operators.  This description captures both the continuous
dissociation of pre-formed bound states and the rare stochastic recombination of
uncorrelated open-charm pairs into bound configurations.
The use of the adiabatic approximation allows us to project the dynamics onto
instantaneous eigenstates of the effective Hamiltonian and to express the
time-dependent survival probability and coalescence rate in analytic form.
In the next Subsections, we apply this framework to derive the non-equilibrium
coalescence probability for the $X(3872)$ in an expanding thermal medium,
establishing the microscopic link between the stochastic evolution of the
density matrix and the observable yield of the bound state at freeze-out.

\subsection{Recombination rate}

The dynamics of a heavy
$c\bar c$ pair immersed in a thermal medium can be formulated through a
Lindblad equation,
\begin{equation}
\partial_t\rho = -i H_{\mathrm{eff}}\rho + i\rho H_{\mathrm{eff}}^\dagger
+ \sum_n C_n \rho C_n^\dagger,
\label{eq:lindblad}
\end{equation}
where $H_{\mathrm{eff}} = H - \tfrac{i}{2}\sum_n C_n^\dagger C_n$
is a non-Hermitian effective Hamiltonian and the $C_n$ are
Lindblad operators encoding stochastic transitions between color
configurations and interaction channels with the medium.

Although $H_{\mathrm{eff}}$ is
non-Hermitian, we can still look for a basis that diagonalizes this Hamiltonian
The eigenfunctions can be
localized bound states or free states. Then, we can define projection operators $P_b$ and $P_f$ onto bound and free
subspaces, respectively, and write the reduced evolution equations,
\begin{align}
\partial_t \rho_b &= -i H_{\mathrm{eff}}\rho_b + i\rho_b H_{\mathrm{eff}}^\dagger\nonumber \\
&\hskip 0.4cm + \sum_n ( P_b C_n\rho_f C_n^\dagger P_b + P_b C_n\rho_b C_n^\dagger P_b ), \\
\partial_t \rho_f &= -i H_{\mathrm{eff}}\rho_f + i\rho_f H_{\mathrm{eff}}^\dagger\nonumber \\
&\hskip 0.4cm 
+ \sum_n ( P_f C_n\rho_f C_n^\dagger P_f + P_f C_n\rho_b C_n^\dagger P_f ).
\end{align}
The Lindblad equation can also be written in an infinitesimal time-step form as
\begin{align}
\nonumber
\rho(t+\mathrm{d}t) =
&\big(1 - \mathrm{Tr}(\Gamma \rho)\,\mathrm{d}t\big)
\frac{(1 - i H_{\mathrm{eff}}\,\mathrm{d}t)\,
\rho(t)\,
(1 + i H_{\mathrm{eff}}^{\dagger}\,\mathrm{d}t)}
{\big(1 - \mathrm{Tr}(\Gamma \rho)\,\mathrm{d}t\big)}\\
&+ \sum_{n}
\mathrm{Tr}(\Gamma_{n}\rho)\,\mathrm{d}t\,
\frac{C_{n}\rho C_{n}^{\dagger}\,\mathrm{d}t}
{\mathrm{Tr}(\Gamma_{n}\rho)\,\mathrm{d}t},
\label{eq:lindblad_discrete}
\end{align}
where 
\nestor{$\Gamma_n = C_n^\dagger C_n$}
defines a partial decay width,
and $\Gamma = \sum_n \Gamma_n$ the total one. 
This representation of the Lindblad equation (which is inspired by the Quantum Trajectories Method~\cite{Dalibard:1992zz}) has a clear probabilistic interpretation:
during a time step $\mathrm{d}t$, the system evolves under the
non-Hermitian Hamiltonian $H_{\mathrm{eff}}$ with probability
$1-\mathrm{Tr}(\Gamma\rho)\,\mathrm{d}t$, or undergoes a quantum jump
associated with operator $C_{n}$ with probability
$\mathrm{Tr}(\Gamma_{n}\rho)\,\mathrm{d}t$.

The probability that a jump from a free state produces a bound state through
channel $n$ is therefore
\begin{equation}
 \mathrm{Tr}\!\left(C_n^{\dagger} P_b C_n \rho_f \right).
\end{equation}
Writing $P_b = \sum_i |i\rangle \langle i|$ in terms of the eigenfunctions of $H_{\mathrm{eff}}$, we define the target density matrix
\begin{equation}
\rho_{t,n} = \frac{C_n^{\dagger} P_b C_n}{Z_{t,n}}, \qquad 
Z_{t,n} = \mathrm{Tr}\!\left(C_n C_n^{\dagger} P_b\right)
  = \sum_i \langle i | C_n C_n^{\dagger} | i \rangle.
\end{equation}
The corresponding transition rate from a free to a bound state through the channel corresponding to the collapse operator $C_n$ is
\begin{equation}
R_n = Z_{t,n}\,\mathrm{Tr}\!\left(\rho_{t,n}\rho_f\right).
\end{equation}
Rewriting this rate in terms of the Wigner distribution allows a closer contact with the traditional coalescence model~\cite{Fries:2008hs}. However, note that the relevant quantity here is the Wigner transform of the target density matrix. This is what we call the target Wigner distribution. We remind that the Wigner transform of a density matrix is,
\begin{equation}
W(\mathbf{R},\mathbf{p}) =
  \int d^3r\, e^{i\mathbf{p}\cdot\mathbf{r}}\,
  \langle \mathbf{R}+\tfrac{\mathbf{r}}{2} | \rho |
          \mathbf{R}-\tfrac{\mathbf{r}}{2} \rangle.
\end{equation}
Then, we obtain
\begin{equation}
R_n = Z_{t,n}
  \int \frac{d^3R\,d^3p}{(2\pi)^3}\,
  W_{t,n}^{\dagger}(\mathbf{R},\mathbf{p})
  W_f(\mathbf{R},\mathbf{p}).
\end{equation}
Assuming that free heavy-quark pairs are uncorrelated and thermalized, their Wigner function reads
\begin{equation}
W_f(\mathbf{R},\mathbf{r},\mathbf{P},\mathbf{p})
  = Z\,N^2 e^{-\frac{\mathbf{P}^2}{4MT} -\frac{ \mathbf{p}^2}{MT}},
  \qquad
  Z^{-1} = V^2 \left(\frac{MT}{2\pi}\right)^3,
\end{equation}
where $M$ is the heavy-quark mass, $V$ the system volume, and \nestor{$N=N_c=N_{\bar c}$ the number of free heavy quarks}.  
Neglecting the center-of-mass dependence of $\rho_{t,n}$, the rate simplifies to
\begin{equation}
R_n = \frac{Z_{t,n} N^2}{ V
  \left(\frac{MT}{4\pi}\right)^{3/2}}
  \int \frac{d^3p}{(2\pi)^3}
  e^{-\frac{\mathbf{p}^2}{MT}}
  W_{t,n}(\mathbf{p}).
  \label{eq:rate}
\end{equation}

We note that in the limit $|\mathbf{p}| \ll \sqrt{MT}$, the exponential approaches unity and we get the simplified expression
\begin{equation}
R_n = \frac{Z_{t,n} N^2}{ V
  \left(\frac{MT}{4\pi}\right)^{3/2}}.
\end{equation}
This implies that, at high temperatures, the recombination rate is not very sensitive to the details of the Wigner distribution of the bound states.

Equation~\eqref{eq:rate} gives the \emph{probability per unit time} for a transition from a free to a bound state through channel $n$.
Now, let us compute the conditional probability that, if this happens at a given time $t$, 
we obtain a particular bound state $i$ at a later time $t_f$. 
To do so, we evolve the density matrix
\begin{equation}
\rho_n =
\frac{P_b C_n \rho_f C_n^{\dagger} P_b}{R_n}.
\end{equation}
In the case in which $\rho_f$ corresponds to thermal equilibrium, we have
\begin{equation}
\rho_n =
\frac{1}{Z_{t,n}
\displaystyle \int \frac{d^3p}{(2\pi)^3} 
e^{-\frac{p^2}{MT}} W_{t,n}(\mathbf{p})}
{\displaystyle \int \frac{d^3p}{(2\pi)^3} e^{-\frac{p^2}{MT}}}
P_b C_n | \mathbf{p} \rangle \langle \mathbf{p} | C_n^{\dagger} P_b .
\end{equation}
Assuming that bound state production after a recombination event is dominated by quantum trajectories (in the sense of the quantum trajectories method) in which no additional quantum jumps happens, the conditional probability that this process produces the bound state 
$|i\rangle$ at time $t_f$ is
\begin{equation}
P(i, t_f | n, t) =
\langle i | e^{-i H_{\mathrm{eff}} (t_f - t)} 
\rho_n(t)
e^{\,i H_{\mathrm{eff}}^{\dagger} (t_f - t)} | i \rangle .
\label{eq:indept}
\end{equation}

In summary, the contribution from coalescence to the probability of observing 
a bound state $i$ at time $t_f$ is given by
\begin{equation}
\int_0^{t_f} dt 
\sum_n R_n(t)\, P(i, t_f | n, t) .
\end{equation}

\subsection{Adiabatic limit}

Let us now consider that the Hamiltonian and the Lindblad operators vary slowly with  time, opposite to what was assumed in Equation~\eqref{eq:indept} .   
Now $R_n(t)$ is time-dependent and the conditional probability is modified to
\begin{equation}
P(i, t_f | n, t) =
\langle i(t_f) | 
e^{-i \int_t^{t_f} dt'\, H_{\mathrm{eff}}(t')}
\rho_n(t)\,
e^{\,i \int_t^{t_f} dt'\, H_{\mathrm{eff}}^{\dagger}(t')}
| i(t_f) \rangle .
\end{equation}
Applying the adiabatic theorem and using the fact that 
$H_{\mathrm{eff}} - H_{\mathrm{eff}}^{\dagger}$ is a $c$-number since the decay width is a constant, the expression simplifies to
\begin{equation}
P(1S, t_f | n, t) =
\langle 1S | \rho_n(t) | 1S \rangle\, S(t_f, t),
\end{equation}
where $S(t_f, t)$ is the survival probability of the bound state between times $t$ and $t_f$.

Therefore, the contribution from coalescence to the probability of obtaining a 
state $1S$ at time $t_f$ is, in the adiabatic approximation,
\begin{equation}
\int_0^{t_f} dt\, S(t_f, t)
\sum_n 
\langle 1S | \rho_n(t) | 1S \rangle\, R_n(t).
\label{eq:SR}
\end{equation}
We can further simplify this by defining the effective transition rate
\begin{equation}
R_{1S}(t) = 
\sum_n \langle 1S | \rho_n(t) | 1S \rangle\, R_n(t)
= 
\langle 1S | 
\sum_n C_n \rho_f C_n^{\dagger}
| 1S \rangle.
\end{equation}
Then, Equation~\eqref{eq:SR} reads
\begin{equation}
\int_0^{t_f} dt\, S(t_f, t) R_{1S}(t).
\end{equation}
Introducing the $1S$ target density matrix
\begin{equation}
\rho_{t,1S} =
\frac{\sum_n C_n^{\dagger} | 1S \rangle \langle  1S | C_n}{Z_{t,1S}},
\ \ 
Z_{t,1S} = 
\langle 1S | \sum_n C_n C_n^{\dagger} | 1S \rangle,
\end{equation}
the recombination rate can be written as
\begin{equation}
R_{1S}(t) = \frac{
Z_{t,1S}(t)\, N^2}{ V
\left(\frac{MT}{4 \pi}\right)^{3/2}}
\int \frac{d^3p}{(2\pi)^3}
e^{-\frac{p^2}{MT}} W_{t,1S}(\mathbf{p}).
\end{equation}
As we mentioned before, we are going to assume that $W_{t,1S}\approx W_{1S}$, where $W_{1S}(p)$ is just the square of the $1S$ wave function in momentum space. This can be computed using the potential discussed in \cite{Armesto:2024zad}.

We now discuss the validity of the adiabatic limit. Since our decay width is constant, the evolution governed by the non-Hermitian Hamiltonian simplifies to that of an Hermitian Hamiltonian multiplied by a c-number. Therefore, standard results for Hermitian systems, including the adiabatic theorem, are directly applicable to our case. As discussed in~\cite{Messiah+1995+583+750}, the adiabatic limit is a good approximation as long as
\begin{equation}
\frac{|\langle i|\frac{dV}{dt}|j\rangle|^2}{(E_i-E_j)^4}\ll 1
\end{equation}
for any two eigenvalues ($i$ and $j$) of the Hamiltonian. We have checked that this is indeed the case. Note that the time dependence of the potential is due to the variation of the temperature with time, as described in the next Subsection.
\subsection{Bjorken-expanding medium}

We now embed this expression in a Bjorken-expanding medium~\cite{Bjorken:1982qr} characterized by
$t T^3 = t_0 T_0^3$, where $(t_0, T_0)$ are the initial proper time and temperature.
Assuming the temperature is homogeneous in the transverse plane, the survival
probability~\cite{Armesto:2024zad} factorizes as
\begin{equation}
S(t_f,t) = S_{-1}(t_f,t) S_2(t_f,t),
\end{equation}
with
\begin{align}
\log S_{-1} &= -\frac{3}{2} \,\nestor{2}A_{-1} T
t
\left[\left(\frac{t_f}{t}\right)^{2/3} - 1\right], \\
\log S_2 &= -\frac{3}{5}\,\frac{\nestor{2}A_2Tt}{m_D^3(T)}
\left[\left(\frac{t_f}{t}\right)^{5/3} - 1\right],
\end{align}
where $A_{-1}$ and $A_2$ are the constants introduced in Ref.~\cite{Armesto:2024zad}  and given in Section~\ref{sec:potential}.
Under these conditions,
\begin{equation}
R_{1S}(t) =
\frac{Z_{t,1S}(T) N^2}{2 A t
\left(\frac{M T}{4\pi}\right)^{3/2}}
\!\int\!\frac{d^3p}{(2\pi)^3} e^{-\frac{p^2}{M T}}
W_{t,1S}(\mathbf{p},T),
\label{eq:Rbjorken}
\end{equation}
where the volume of the medium has been approximated by the area of the transverse overlap
of the colliding nuclei times the longitudinal length of the medium.

\subsubsection{Low-temperature limit}

For $\sqrt{MT} \ll p_{\mathrm{bs}}$, where $p_{\mathrm{bs}}$ is the characteristic momentum scale
of the bound-state wavefunction, the integral in Equation~\eqref{eq:Rbjorken}
reduces to
\begin{align}
\nonumber
\int\!\frac{d^3p}{(2\pi)^3} e^{-\frac{p^2}{M T}} W_{t,1S}(\mathbf{p},T)
&\simeq W_{t,1S}(\mathbf{0},T)
\!\int\!\frac{d^3p}{(2\pi)^3} e^{-\frac{p^2}{M T}}\nonumber \\
&=W_{t,1S}(\mathbf{0},T)
\!\left(\frac{M T}{4\pi}\right)^{3/2}.
\end{align}
Note that $W_{t,1S}(\mathbf{0},T) = |\psi_{1S}(\mathbf{p}=\mathbf{0})|^2$.
Assuming a Gaussian form for the bound-state wavefunction,
\begin{equation}
\psi_{1S}(\mathbf{r}) =
\left(\frac{3}{2\pi \langle r^2\rangle}\right)^{3/4}
\exp\!\left[-\frac{3 r^2}{4\langle r^2\rangle}\right],
\label{eq:gaussian}
\end{equation}
we find
\begin{equation}
\psi_{1S}(\mathbf{p}=\mathbf{0}) = \int d^3r\, \psi_{1S}(\mathbf{r})
= \left(\frac{8\pi \langle r^2\rangle}{3}\right)^{3/4}.
\end{equation}
Therefore, in this limit, the coalescence rate simplifies to
\begin{equation}
R_{1S}(t) =
Z_{t,1S}(T) N^2
\left(\frac{8\pi \langle r^2\rangle}{3}\right)^{3/2}
\frac{1}{2 A t}.
\label{eq:R_lowT}
\end{equation}

We observe that in this limit the recombination rate depends on a geometrical factor proportional to the volume occupied by the bound state divided by the volume of the medium. However, this approximation never holds for the $X(3872)$ at nowadays heavy-ion experiments since the typical value of $\sqrt{MT}$ is too large.

\subsection{Phenomenological inputs}

To obtain quantitative estimates for the coalescence contribution to the $X(3872)$ yield, 
we must specify several phenomenological inputs entering Equation~\eqref{eq:Rbjorken} 
and the kinetic evolution. These inputs determine the effective volume, initial charm 
production, and reference normalization for the nuclear modification factor.

\begin{itemize}
    \item Medium volume:
    The effective volume of the fireball is approximated by the transverse overlap area of 
    the colliding nuclei times the longitudinal extension of the medium. The transverse area 
    is computed from a standard Glauber model supplemented with the shadowing model used in \cite{Escobedo:2021ifp}. We define the transverse area as the area in which the initial temperature is larger than the freeze-out temperature. The longitudinal length is taken as $L = 2 t$, with $t$ the proper time at which the coalescence process occurs.

    \item Initial open-charm production: 
    The initial number of unbound charm quarks is estimated from binary nucleon-nucleon 
    scaling
        $N_{c\bar{c}}^{(0)} = N_{\mathrm{coll}} \frac{\sigma_{pp\to c\bar{c}}}{\sigma_{pp}}$,
    where $N_{\mathrm{coll}}$ is the number of binary collisions obtained from the Glauber 
    model, $\sigma_{pp} = 70\,\mathrm{mb}$ is the total inelastic proton-proton cross section, 
    and $\sigma_{pp\to c\bar{c}}$ is taken from Ref.~\cite{ALICE:2021dhb}. We use the midrapidity value appropriate for $\sqrt{s_{NN}}=5.02~\mathrm{TeV}$, $d\sigma_{pp\to c\bar{c}}/dy=1.165$ mb, which is the $p_T$-integrated value down to 0.

    \item Initial $X(3872)$ production:
    To compute the nuclear modification factor $R_{AA}$, we normalize the final yield 
    to the number of $X(3872)$ that would be produced in the absence of medium effects,
   $N_{\mathrm{coll}}\,\frac{\sigma_{pp\to X}}{\sigma_{pp}}$.
    The cross section $\sigma_{pp\to X}$ is taken from the LHCb measurement 
    of prompt $X(3872)$ production~\cite{LHCb:2011zzp}, corrected for the branching 
    ratio ${\cal B}(X(3872)\!\to\! J/\psi\pi^+\pi^-)$ from the PDG~\cite{ParticleDataGroup:2024cfk} and adapted at $\sqrt{s_{NN}}=5.02~\mathrm{TeV}$. The resulting effective cross section 
    is $\sigma_{pp\to X} = 0.126\,\mu\mathrm{b}$
    measured in the rapidity range 2.5-4.5 and transverse momentum in the range 5-20 GeV/c.
    As for the $\sigma_{pp\to c\bar{c}}$, we are interested in midrapidity results. In order to take into account the difference
    in rapidity coverage, we apply a factor 1.58, assuming that the increase at midrapidity will follow the $J/\psi$ and the open charm behavior~\cite{ALICE:2021dhb,LHCb:2016ikn,ALICE:2019pid,ALICE:2021qlw,Wu:2024gil}.

    \item 
    The above experimental result for the $X(3872)$ cross section is obtained in the $p_T$-range 5-20 GeV/c. As for the $\sigma_{pp\to c\bar{c}}$, we are interested in $p_T$-integrated values, so we correct using the $p_T$ distribution developed in Ref.~\cite{Bossu:2011qe}. The corresponding integrated value exceeds by a factor of 14 the one obtained in the $p_T$-range 5-20 GeV/c.

    \item Finally, and for simplicity, we assume that $Z_{t,1S}$ is equal to the decay width of the 1S state. The reason is the following. Due to the fluctuation-dissipation theorem, the transition rate from the 1S state to another state $\ell$ ($\Gamma_{1S\to\ell}$) is equal to the rate of the opposite transition ($Z_{t,\ell\to 1S}$) multiplied by $\mathrm{e}^{\frac{E_{1S}-E_\ell}{T}}$. On the other hand, for any species $\phi,\psi$ which can  transition one into each other,
    $$\Gamma_\phi=\sum_{\psi}\Gamma_{\phi\to\psi}=\sum_{\psi}|\langle\psi|C_n|\phi\rangle|^2,$$
$$Z_\phi=\sum_{\psi}\Gamma_{\psi\to\phi}=\sum_{\psi}|\langle\phi|C_n|\psi\rangle|^2.$$
Since in our approach $E_{1S}-E_\ell< E_g\ll T$, with $E_g$ the binding energy of the state -- see the discussion above Equation~\eqref{eq:LindbladOrig}, we take for simplicity the exponential to be one. This is equivalent to consider that $C_n^\dagger=C_n$. It is a well-known result that Lindblad equations with this property increase entropy monotonically~\cite{Benatti:1987dz}. 

\end{itemize}

These phenomenological parameters fix the initial conditions and normalization of 
Equation~\eqref{eq:Rbjorken}, allowing for a comparison 
between the model predictions and the measured $R_{AA}$ of the $X(3872)$.

\subsection{Results}
\label{sec:results}

Using the phenomenological inputs described in the previous Section and the parameters given below Equation~\eqref{eq:survival}, we compute the time-integrated yield of the $X(3872)$ from Equation~\eqref{eq:Rbjorken}:
\begin{equation}
N_X(t_f) = \int_0^{t_f} dt\, S(t_f, t)\, R_{1S}(t),
\label{eq:Nxevo}
\end{equation}
where $R_{1S}(t)$ is the recombination rate and $S(t_f,t)$ the survival probability of the bound state. Note that Equation~\eqref{eq:Rbjorken} assumes an homogeneous temperature in the medium. However, it is straight-forward to relax this approximation to include the transverse position dependence of the initial temperature by applying the previous formula to each differential of area, $dS$. For simplicity, we consider that the density of unbound heavy quarks is
\begin{equation}
    W_c(\mathbf{p},\mathbf{r})=\frac{N_c}{\left(\frac{MT(\mathbf{r})}{2\pi}\right)^{3/2}V}e^{-\frac{p^2}{2MT(\mathbf{r})}}\ ,
\end{equation}
where $\mathbf{p}$ and $\mathbf{r}$ are the momentum and position of the unbound heavy quark, respectively. This could be improved by simulating the behavior of the unbound quarks distribution with a Fokker-Planck equation, but we leave this to the future.

A crucial input is the Wigner density of the bound state. This is obtained using the potential model that we developed in \cite{Armesto:2024zad}. We solve the corresponding Schrödinger equation to obtain the bound state wave function. Using the definition of the Wigner transform, we can obtain the Wigner density from the wave function. 

\begin{figure}[ht]
    \centering
    \includegraphics[width=0.48\textwidth]{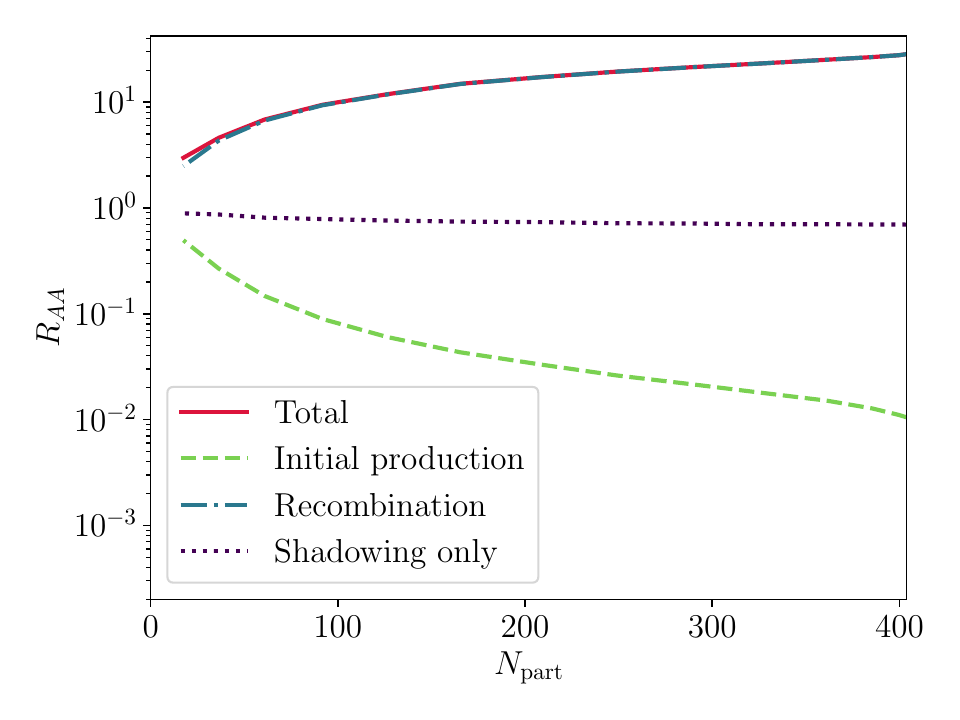}
    \caption{Midrapidity $p_T$-integrated $R_{AA}$ for $X(3872)$  for Pb-Pb collisions at $\sqrt{s_{NN}}=5.02~\mathrm{TeV}$ including suppression and recombination. The dotted lines represents the effect of shadowing modeled as in \cite{Escobedo:2021ifp}, which is included in all next scenarios. The dashed line corresponds to the suppression of the $X(3872)$ states that are initially present in the plasma. The contribution from recombined pairs is represented by the dotted-dashed line. Finally, the solid line represents the full contribution to $R_{AA}$. }
    \label{fig:Raarecom}
\end{figure}

Figure~\ref{fig:Raarecom} shows\footnote{While the increasing trend of the recombination result is generic, the concrete shape depends on the modeling of the number of charm quarks versus the number of participants, which relies on the corresponding functional forms of the number of binary nucleon-nucleon collisions and of the shadowing~\cite{Escobedo:2021ifp}.} the resulting nuclear modification factor $R_{AA}$, including both suppression and coalescence contributions, as a function of the number of participants $N_{\mathrm{part}}$.
Note that our predicted value is integrated over all transverse momenta, thus dominated by low $p_T$ values. In contrast, the experimental measurement corresponds to the high transverse momentum range $15 < p_T < 50 ~\mathrm{GeV}$, where recombination, if it is at play, is expected to be much weaker than in the low-$p_T$ range. Therefore, at low transverse momenta where coalescence processes are expected to dominate, the regeneration contribution should be comparatively larger and the nuclear modification factor could amount to several units.
Although our model is quite simple and qualitative, we are encouraged by the fact that our results \nestor{look} compatible with the available data when the different $p_T$ ranges are taken into account. 
Moreover, let us note that our main motivation is to understand if $R_{AA}$ is qualitatively different for tetraquarks and molecular states. To arrive to a definitive conclusion regarding this, future studies applying the formalism that we have developed in this manuscript and in \cite{Armesto:2024zad} to the molecular case are needed.
\elena{Still, we would like to emphasise that the present experimental uncertainties are too large to draw firm conclusions. Only future measurements with improved precision and extended coverage towards lower $p_\perp$ compared to the corresponding $p_\perp$-differential model predictions will provide a stringent test of the predicted production mechanism. Such a $p_\perp$-differential calculation would require modeling the momentum spectrum of thermalized charm quarks and its mapping onto the recombined bound state, an extension that lies beyond the qualitative, order-of-magnitude scope of the present work and that we leave for future study.}

The above results were obtained using numerical simulations. We find that analytical approximations might be useful to qualitatively understand our results. Equation~\eqref{eq:Nxevo} fulfills the following differential equation:
\begin{equation}
    \frac{dN_X}{dt_f}=-\Gamma_{1S}(t_f)N_X(t_f)+R_{1S}(t_f)\,.
    \label{eq:difeq}
\end{equation}
In the simple case in which $\Gamma_{1S}$ and $R_{1S}$ are constants, $N_X$ tends to the asymptotic value $\frac{R_{1S}}{\Gamma}$. This is a good approximation to the final number of $X$ given by Equation~\eqref{eq:Nxevo} if the medium temperature changes very slowly. Following this argument, we would estimate the total number of $X$ particles created in a heavy-ion collision as 
\begin{equation}
    N_X\sim \frac{
 N^2}{ V
\left(\frac{MT}{4\pi}\right)^{3/2}}
\int \frac{d^3p}{(2\pi)^3}
e^{-\frac{p^2}{MT}} W_{t,1S}(\mathbf{p}).
\label{eq:numNX}
\end{equation}
Since the integral is, by definition, a number between $0$ and $1$, the order of magnitude of $N_X$ will be simply given by $\frac{
 N^2}{ V
\left(\frac{MT}{4\pi}\right)^{3/2}}$.
Therefore, the contribution of regeneration to the nuclear modification factor, $R_{AA}$, is
\begin{equation}
    R_{AA,recomb}\sim\frac{N^2}{V
\left(\frac{MT}{4\pi}\right)^{3/2}N_{X,T=0}}\,,
\end{equation}
where $N_{X,T=0}$ is the number of $X$ states that would be produced in a heavy-ion collision if there were no medium effects.  Using the model in \cite{Escobedo:2021ifp}, $N=N_{col}S^{sh}\frac{\sigma_{c\bar{c}}}{\sigma_{pp}}$ where $N_{col}$ is the number of binary collisions in the Glauber model, $S^{sh}$ is the shadowing factor, $\sigma_{pp}$ is the inelastic proton-proton cross section and $\sigma_{c\bar{c}}$ is the production cross section of a $c\bar{c}$ pair in a $pp$ collision. On the other hand, $N_{X,T=0}=N_{col}\frac{\sigma_X}{\sigma_{pp}}$. This implies that
\begin{equation}
    R_{AA,recomb}\sim\frac{N_{col}(S^{sh})^2\sigma_{c\bar{c}}^2}{\sigma_{pp}\sigma_XV
\left(\frac{MT}{4\pi}\right)^{3/2}}\,.
\label{eq:estimate}
\end{equation}
As an illustration, let us look at central collisions ($N_{col}\sim 1500$) and assume that the steady state is reached at a freeze-out time of around $\tau \sim 15.6\,\textrm{fm}$, when the temperature is around $T\sim 200\,\textrm{MeV}$. Then, the previous equation gives a value of $R_{AA,recomb}\sim 10$, of the same order of magnitude as we find in our numerics.

\subsection{Considering the feedback on \texorpdfstring{$N_c$}{Nc}}

The results discussed above indicate a quite large increase of $R_{AA}$. To double check this conclusion, we will now estimate the effect of relaxing one of our main assumptions. Namely, that the formation of bound states does not substantially reduce the amount of unbound $c\bar{c}$ pairs in the medium. Let us note that we expect that the main effect will come from the feedback onto $N_c$ of the recombination of conventional quarkonium states. 
This effect can be studied by a slight modification of Equation~\eqref{eq:difeq}. 
The time evolution of the $J/\psi$ and $X(3872)$ abundances is described by
\begin{align}
&\frac{dN_{J/\psi}}{dt}
= -\Gamma_{J/\psi}\,N_{J/\psi}
\label{eq:feedback_Jpsi}  \\ &+Z_{t,J/\psi}(T)\,
    \frac{(N_0 - N_{J/\psi} - N_X)^2}{2A\,t
    \left(\frac{MT}{4\pi}\right)^{3/2}}
    \int\!\frac{d^3p}{(2\pi)^3}\,
    e^{-\frac{p^2}{MT}}\,
    W_{t,1S}(\mathbf{p},T),
    \nonumber\\
&\frac{dN_X}{dt}
= -\Gamma_X\,N_X\nonumber
  \\ &+Z_{t,X}(T)\,
    \frac{(N_0 - N_{J/\psi} - N_X)^2}{2A\,t
    \left(\frac{MT}{4\pi}\right)^{3/2}}
    \int\!\frac{d^3p}{(2\pi)^3}\,
    e^{-\frac{p^2}{MT}}\,
    W_{t,X}(\mathbf{p},T),
\nonumber
\end{align}
where $N_0$ denotes the initial number of $c\bar{c}$ pairs, 
$\Gamma_i$ are the in-medium dissociation widths, 
and $Z_{t,i}(T)$ and $W_{t,i}(p,T)$ encode the microscopic structure of the corresponding 
bound states.  
These equations account for the mutual depletion of the unbound-charm reservoir by the 
formation of both $J/\psi$ and $X(3872)$, ensuring particle-number conservation within the 
Lindblad-based kinetic framework.

In the case of the $X(3872)$, the effective non-Hermitian Hamiltonian can be approximated 
as $H_{\mathrm{eff}} = H - i\Gamma/2$, with a momentum-independent width $\Gamma$, 
which allows the direct use of the adiabatic theorem in the time evolution. 
For the $J/\psi$, however, the imaginary part of the potential depends on the relative 
momentum, rendering $\Gamma$ an operator rather than a constant. 
In this case, a generalized adiabatic treatment must be employed, following the formalism 
of Ref.~\cite{Kumar:2025quk}. 

However, looking at the structure of the couple differential equations we can already estimate that the feedback on $N_c$ is small. Let us consider that $N_0\gg N_{J/\Psi}\gg N_X$. Then
\begin{equation}
\begin{split}
    \frac{dN_{J/\psi}}{dt}
= -\Gamma'_{J/\psi}\,N_{J/\psi}\\
+Z_{t,J/\psi}(T)\,
    \frac{(N_0)^2}{2A\,t
    \left(\frac{MT}{4\pi}\right)^{3/2}}
    \int\!\frac{d^3p}{(2\pi)^3}\,
    e^{-\frac{p^2}{MT}}\,
    W_{t,1S}(\mathbf{p},T)
+\mathcal{O}\left(\frac{N_X}{N_0}\right),
\end{split}
\end{equation}
where
\begin{equation}
    \Gamma'_{J/\psi}=\Gamma_{J/\psi}+\frac{Z_{t,J/\Psi}(T)N_0}{At\left(\frac{MT}{4\pi}\right)^{3/2}}\int\!\frac{d^3p}{(2\pi)^3}\,
    e^{-\frac{p^2}{MT}}\,
    W_{t,1S}(\mathbf{p},T)
+\mathcal{O}\left(\frac{N_X}{N_0}\right).
\end{equation}
Therefore, we can check if the backreaction effect is important by looking at the correction to the decay width,
\begin{equation}
    \Gamma'_{J/\psi}=\Gamma_{J/\psi}+Z_{t,J/\Psi}(T)\gamma\int\!\frac{d^3p}{(2\pi)^3}\,
    e^{-\frac{p^2}{MT}}\,
    W_{t,1S}(\mathbf{p},T)
+\mathcal{O}\left(\frac{N_X}{N_0}\right)\,,
\end{equation}
where the parameter $\gamma$ is defined as
\begin{equation}
    \gamma=\frac{2N_{col}\sigma_{c\bar{c}}}{\sigma_{pp}V\left(\frac{MT}{4\pi}\right)^{3/2}}\,.
    \label{eq:estgamma}
\end{equation}
We obtain for this parameter values smaller than $0.16$ in the most unfavorable conditions (those present at the beginning of the medium evolution when the temperature is higher and the medium volume smaller). In conclusion, it is a good approximation to consider that the number of unbound charm quarks in unaffected by the formation of bound states.

\section{Chemical equilibrium coalescence}
\label{sec:chemical}

{ In this Section we discuss a different but related model.} In this case, we consider a thermalized system composed of three particle species: heavy quarks $Q$ and antiquarks $\bar{Q}$, a conventional quarkonium  bound state species $\Pi$ and an exotic bound state $\chi$. The reactions
\begin{equation}
Q + \bar{Q} \leftrightarrow \Pi\,,
\end{equation}
and 
\begin{equation}
    Q+\bar{Q}\leftrightarrow\chi\,.
\end{equation}
establish chemical equilibrium among these constituents, characterized by the chemical potentials 
$\mu_Q = \mu_{\bar{Q}} = \mu$, $\mu_{\Pi} = 2\mu$ and $\mu_\chi=2\mu$. 
The energy gap between two unbound heavy quarks and the bound $\Pi$ state is denoted by $E_g$ and that between two unbound heavy quarks and the exotic bound $\chi$ state is $E^e_g$. Within the grand canonical ensemble, the probability of a macroscopic configuration $n$ is given by
\begin{equation}
P_n = \exp\!\left[\frac{\Omega + \mu_Q N_Q^n + \mu_{\bar{Q}} N_{\bar{Q}}^n + \mu_{\Pi} N_{\Pi}^n+\mu_\chi N_\chi^n - E_n}{T}\right],
\end{equation}
where $E_n$ is the total energy of the state and $\Omega$ is the grand potential, defined through the normalization condition $\sum_n P_n = 1$. Substituting the equilibrium conditions for the chemical potentials yields
\begin{equation}
\Omega = -T \log \!\left[\sum_n \exp\!\left(\frac{\mu(N_Q^n + N_{\bar{Q}}^n + 2N_{\Pi}^n+2N^n_\chi) - E_n}{T}\right)\right].
\end{equation}

The contributions to $\Omega$ from the individual species are computed separately. For heavy quarks, which are fermions with two spin states, the result in the continuum limit is
\begin{equation}
\Omega_Q = -2T V \int \frac{d^3p}{\pi^3} 
\log\!\left(1 + e^{(\mu - \frac{p^2}{2M})/T}\right),
\end{equation}
and an identical expression holds for $\bar{Q}$. 
For the $\Pi$ meson, assuming a boson with three polarization states and a scalar partner like in the case of the $(J/\Psi,\eta_c)$ multiplet, the contribution becomes
\begin{equation}
\Omega_{\Pi} = 4T V \int \frac{d^3p}{\pi^3} 
\log\!\left(1 - e^{(2\mu + E_g - \frac{p^2}{4M})/T}\right)\,.
\end{equation}
Finally, for the $\chi$ meson, 
\begin{equation}
\Omega_{\chi} = 3T V \int \frac{d^3p}{\pi^3} 
\log\!\left(1 - e^{(2\mu + E^e_g - \frac{p^2}{4M})/T}\right)\,.
\end{equation}
{ Note that we assume a spin degeneracy factor $g_\chi=3$ for the $\chi$ family, corresponding to the three polarization states of the $X(3872)$ ($J=1$). In the statistical model, the $X(3872)$ is treated as an effective point-like hadronic degree of freedom, and its internal structure is not explicitly taken into account. This is an approximation. However, varying the multiplicity of the $\chi$ states within reasonable limits does not qualitatively affect our conclusions. } The total grand  potential of the system reads
\begin{align}
\nonumber
\Omega &= -4T V \int \frac{d^3p}{\pi^3} 
\log\!\left(1 + e^{(\mu - \frac{p^2}{2M})/T}\right)
\\ &+4T V \int \frac{d^3p}{\pi^3} 
\log\!\left(1 - e^{(2\mu + E_g - \frac{p^2}{4M})/T}\right)\nonumber \\
&+3T V \int \frac{d^3p}{\pi^3} 
\log\!\left(1 - e^{(2\mu + E^e_g - \frac{p^2}{4M})/T}\right)\,.
\end{align}

The definition of the grand potential implies the identity
\begin{equation}
\frac{\partial \Omega}{\partial \mu}
= -\big\langle N_Q + N_{\bar{Q}} + 2N_{\Pi}+2N_\chi \big\rangle .
\end{equation}

Hence, given the initial total  number of open and hidden heavy-quark states, the chemical potential is obtained by solving
\[
\frac{\partial \Omega}{\partial \mu}(\mu) + \big\langle N_Q + N_{\bar{Q}} + 2N_{\Pi}+2N_\chi\big\rangle \;=\; 0.
\]

In practice, one proceeds as follows:
\begin{enumerate}
  \item Solve the equation \(\dfrac{\partial \Omega}{\partial \mu}(\mu) + \langle N_Q + N_{\bar{Q}} + 2N_{\Pi}+2N_\chi\rangle = 0\) to obtain \(\mu\). It is often more convenient to solve for the fugacity \(z=\mathrm{e}^{\mu/T}\).
  \item Once \(\mu\) (or \(z\)) is determined, compute the individual equilibrium expectation values via
  \[
  \langle N_Q \rangle = -\frac{\partial \Omega}{\partial \mu_Q}\Bigg|_{\mu_Q=\mu}.\]
\end{enumerate}

To simplify the procedure, we make the \textit{ansatz} that $z$ is small. Later, we will verify that this is indeed the case. Using this approximation,
\begin{align}
    \frac{\partial\Omega}{\partial\mu}=\frac{z}{T}\frac{\partial\Omega}{\partial z}&=2\frac{z}{T}\frac{\partial\Omega_Q}{\partial z}+\mathcal{O}(z^2)\nonumber \\&=-2^{3/2}4zV\left(\frac{MT}{\pi}\right)^{3/2}+\mathcal{O}(z^2)\,.
    \end{align}
This implies that
\begin{equation}
    z=\frac{\big\langle N_Q + N_{\bar{Q}} + 2N_{\Pi}+2N_\chi\big\rangle}{2^{3/2}4V\left(\frac{MT}{\pi}\right)^{3/2}}\,.
\end{equation}
We can check that the fugacity has the same parametric form as the parameter $\gamma$ estimated in Equation~\eqref{eq:estgamma}, but the numerical prefactor makes it much smaller. It is also worth noting that, in the limit of small $z$, the value of the fugacity is not sensitive to the number of bound states found in the medium. In other words, the assumption that there is no backreaction on the number of unbound heavy quarks from the formation of bound states implies the small fugacity limit. But the reverse is not true: small fugacity does not imply small backreaction, which justifies the lack of approach to equilibrium that we observe in our results.

Now, let us look at the number of $\chi$ particles predicted by this model:
\begin{equation}
\begin{split}
    \big\langle N_\chi\big\rangle=-\left.\frac{\partial\Omega_\chi}{\partial\mu_\chi}\right|_{\mu_\chi=2\mu}=24Vz^2\mathrm{e}^{E^e_g/T}\left(\frac{MT}{\pi}\right)^{3/2}+\mathcal{O}(z^4)\\
    =\frac{3\mathrm{e}^{E^e_g/T}\big\langle N_Q + N_{\bar{Q}}\big\rangle^2}{16V\left(\frac{MT}{\pi}\right)^{3/2}}+\mathcal{O}(z^4)\,.
\end{split}
\label{eq:finalchemeq}
\end{equation}
Comparing this with Equation~\eqref{eq:numNX}, we observe that both models lead to a number of exotic states of the same order of magnitude as long as $\mathrm{e}^{E^e_g/T}\sim 1$. This is not at all incoherent since the applicability of the Lindblad equation implies that the energy gap is smaller than the temperature. { However, we cannot regard this as an independent validation as both models are based on the premise that $T\gg E$.} On the other hand, the numerical prefactor makes the result in Equation~\eqref{eq:finalchemeq} sizably smaller than that in Equation~\eqref{eq:numNX}, which aligns with the previous discussion: our results are far from chemical equilibrium. This is in agreement with the observation that the fireball lifetime in heavy ion collisions can be smaller than the quarkonium equilibration time~\cite{Alund:2020ctu,Brambilla:2025sis}.

We also note that this chemical equilibrium model only knows about the spectroscopy of the bound state and, by definition, cannot distinguish a tetraquark from a molecular state. The difference between the two scenarios, regarding recombination, lies in how fast they relax to the equilibrium state. Note that the medium expands in a heavy-ion collision. Therefore, if the expansion is much faster than the relaxation, chemical equilibrium might never be reached.

\section{The Gaussian wave function approximation}
\label{sec:gaussian}
As we have mentioned previously, a dedicated study is needed in order to extend our model to the hadronic molecule case. However, we can get a qualitative approximate understanding by assuming that the wave function of the bound state is Gaussian, like in Equation~\eqref{eq:gaussian}. In this case, 
\begin{equation}
    \int\!\frac{d^3p}{(2\pi)^3}\,
    e^{-\frac{p^2}{MT}}\,
    W_{t,X}(\mathbf{p},T)=\frac{1}{\left(1+\frac{3}{2 MT\langle r^2\rangle}\right)^{3/2}}\,.
\end{equation}
This is a common approximation within the coalescence framework. Typical values of $\langle r^2\rangle$ for a tetraquark state in the vacuum are expected to be in the range $(0.5\,\textrm{fm})^2$ to $(1\,\textrm{fm})^2$ while for a molecule we expect much higher values. 

\begin{figure}[hbt]
    \centering
    \includegraphics[width=0.48\textwidth]{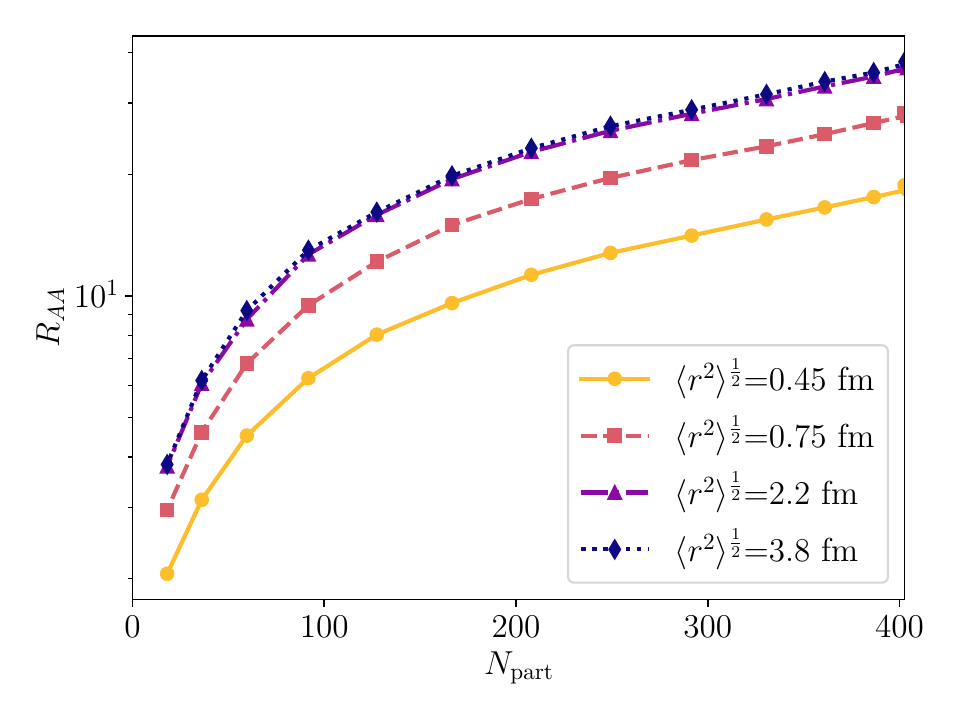}
    \caption{Nuclear modification factor assuming a Gaussian wave function unaffected by the medium, for different values of $\langle r^2\rangle^{\frac{1}{2}}$.}
    \label{fig:gaussian}
\end{figure}

\begin{figure}[hbt]
    \centering
    \includegraphics[width=0.48\textwidth]{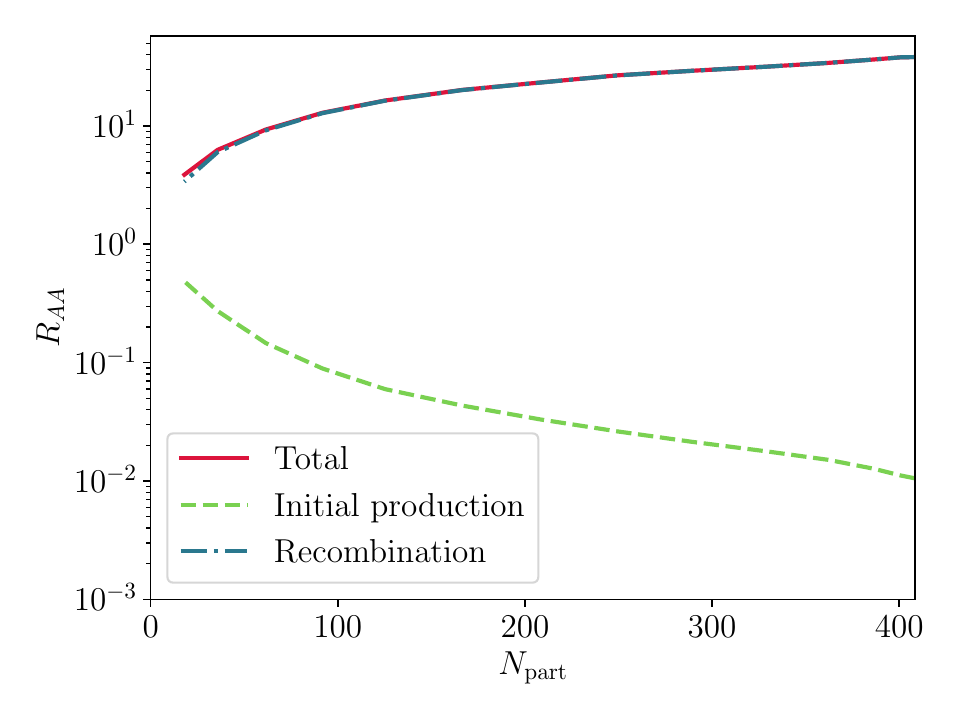}
    \caption{Nuclear modification factor in the limit $\sqrt{MT}\gg p_{bs}$, with $p_{bs}$ the momentum of the bound state.}
    \label{fig:Raa_high}
\end{figure}

In Figure~\ref{fig:gaussian} we show the result of computing the nuclear modification factor assuming that the wave function of the bound state is Gaussian. Note that, differently to the results in Subsection~\ref{sec:results}, here we do not take into account how the temperature affects the wave function, and the difference in decay rate and in dissociation temperature between a tetraquark and a molecule. On general grounds, we expect states to become wider as they get closer to their dissociation temperature. We observe that regeneration is significantly higher for values of $\langle r^2\rangle$ typical of a hadronic molecule compared to that of a tetraquark. In fact, for a molecule, recombination is very close to the high temperature limit discussed below Equation~\eqref{eq:rate}. In Figure~\ref{fig:Raa_high} we show the results in this limit for comparison. This Figure also shows that the effect of the Wigner distribution at finite temperatures is a decrease of the $R_{AA}$ by a few 10\%'s.

\victor{Figure~\ref{fig:gaussian} isolates the size dependence of the recombination rate at fixed thermal history and is not a controlled molecule vs. tetraquark comparison, for which the molecular wave function, dissociation temperature and width would also be needed. Their net effect is not clear a priori.}

Our results point out that it might indeed be possible to distinguish a tetraquark from a hadronic molecule state by studying its nuclear modification factor. However, note that \nestor{our simplified Gaussian approach neglects several effects that may differ the molecular and the tetraquark cases: medium-induced modifications of the molecular wave function, dissociation temperatures, thermal widths and survival probabilities. Therefore, }
a dedicated study of the molecular case would be needed to put \nestor{these} results on \nestor{firm} grounds. 

\section{Summary and conclusions}
\label{sec:summary}

In this work, we have investigated the production of the exotic meson $X(3872)$ in heavy-ion collisions at the LHC, working under the hypothesis that it is a compact tetraquark state. To describe its dynamics in the Quark-Gluon Plasma, we derived a coalescence model based on the Lindblad equation for open quantum systems. This framework allows a consistent treatment of both the dissociation of the initially produced states and the regeneration of the $X(3872)$ from the recombination of thermalized heavy quarks in the medium.

Our main results can be summarized as follows:

\begin{itemize}
    \item \underline{Dominance of Recombination:} We found that if the $X(3872)$ is a compact tetraquark, its production in heavy-ion collisions is dominated by recombination mechanisms rather than by the survival of primordial states. This result holds under the assumption that the unbound heavy quarks are thermalized and that the adiabatic approximation is valid for the evolution of the bound state.

    \item \underline{Enhancement of the Nuclear Modification Factor:} The inclusion of coalescence leads to a significant enhancement of the nuclear modification factor, with predictions for $R_{AA}$ reaching values sizably greater than unity
    in the kinematic region probed by the CMS Collaboration. This is in stark contrast to the strong suppression expected in scenarios neglecting regeneration and\nestor{, considering the different kinematic windows,} it aligns qualitatively with the experimentally observed enhancement of the $X(3872)$ yield relative to the $\psi(2S)$.

    \item \underline{Robustness of the Prediction:} To validate our kinetic approach, we compared our results with a static chemical equilibrium model \nestor{relying on the same hierarchy of scales}. While the chemical equilibrium limit predicts a  smaller recombination contribution due to the cooling of the expanding medium, it confirms that regeneration is the primary source of the final yield and provides a parametric consistency check for our coalescence formalism.

    \item \underline{Sensitivity to Internal Structure:} An alternative exploratory study using a \nestor{simplistic} Gaussian wave function approximation suggests that the recombination rate is highly sensitive to the spatial extent of the state until it becomes very large. Specifically, larger sizes typical of hadronic molecules may lead to even higher regeneration rates than those predicted for the compact tetraquark\victor{, at fixed thermal history; a full comparison would also need the molecular dissociation temperature and width, whose direction is not clear.}
\end{itemize}

In conclusion, our study provides arguments to identify the nuclear modification factor $R_{AA}$ as a \victor{promising} observable for discriminating between the compact tetraquark and hadronic molecule scenarios for the $X(3872)$. The substantial enhancement predicted for the tetraquark case provides a clear signature that can be tested against future high-precision data. Future theoretical work should aim to perform a rigorous calculation of the molecular scenario within the same open quantum system framework to provide a direct quantitative comparison.

Several avenues for future research could further refine the precision of these predictions. First, regarding the medium evolution and heavy quark dynamics, our current framework assumes a simplified Bjorken evolution and fully thermalized heavy quarks. A more rigorous approach would involve coupling our open quantum system formalism to a state-of-the-art hydrodynamical description of the plasma. Furthermore, instead of assuming instantaneous thermalization, the phase-space evolution of unbound heavy quarks could be explicitly studied using some model developed for the study of open heavy flavor diffusion \cite{Rapp:2018qla}.

As we mentioned previously, we assume that the target Wigner distribution is equal to the Wigner distribution of the bound state. If we would know the value of the collapse operators $C_n$, then we would be able to explicitly compute the target Wigner distribution. We do not expect that using a more realistic target Wigner distribution changes qualitatively our results. However, it would be interesting to check them in the future if a more elaborated picture of the exotic hadron-medium interaction is developed.

On the theoretical side of the bound state description, the adiabatic approximation employed here could be relaxed. While solving the full time-dependent Schrödinger equation without this approximation would increase the computational cost, it would provide a robust check on the validity of the adiabatic limit in this context. Additionally, a more comprehensive description of the $X(3872)$ internal structure could be achieved by adopting a hybrid model as proposed in \cite{Braaten:2024tbm,Brambilla:2024imu}. This approach unifies the compact and molecular pictures by employing a potential that behaves like a tetraquark state at short distances and transitions to a hadronic molecule-like interaction at large separations. Implementing such a framework would allow us to dynamically probe the interplay between the compact core and the molecular tail during the evolution in the medium. 

Additionally, the extension of the treatment of the adiabatic approximation to non-Hermitian Hamiltonians would be demanded for the application of the method developed here to the $J/\psi$ -- another avenue for future studies.

Finally, the quantitative accuracy of our results relies heavily on the input parameters derived from proton-proton collisions. Improved experimental measurements of the $c\bar{c}$ pair production cross sections and the $X(3872)$ baseline yields in $pp$ collisions would significantly reduce the systematic uncertainties in our model, leading to more definitive predictions for the nuclear modification factor.

\section*{Acknowledgements}
NA, EGF and VLP are supported by European Research Council project ERC-2018-ADG-835105 YoctoLHC, by Xunta de Galicia (CIGUS Network of Research Centres), by European Union ERDF, and by the Spanish Research State Agency under projects PID20231527\-62NB---I00 and CEX2023-001318-M financed by MICIU/AEI/10.13039/501100011033. The work of MAE has been supported by the Maria de Maetzu excellence program under project CEX2024-001451-M, and by project PID2022-136224NB-C21 funded by MICIU/AEI/10.13039/501100011033, and by grant 2021-SGR-249 of Generalitat de Catalunya. VLP has been supported by Xunta de Galicia under project ED481A2022/286.

\bibliographystyle{elsarticle-num} 
\bibliography{xrecombination}

\end{document}